\documentclass[12pt]{article}

\usepackage{multirow}
\usepackage[table,xcdraw,booktabs]{xcolor}
\usepackage{graphicx}
\usepackage{caption}
\usepackage{times}
\usepackage{booktabs}
\usepackage{amsfonts}
\usepackage{titlesec}
\usepackage{ulem}
\usepackage{natbib}
\usepackage{makecell}
\usepackage{hyperref}  
\usepackage{url}  
\usepackage{indentfirst}

\begin{document}

%
%


\title{SeisCLIP: A seismology foundation model pre-trained by multi-modal data for multi-purpose seismic feature extraction}

%
%
%
%
%
%

\author
{Xu Si$^{1}$, Xinming Wu$^{1\ast}$, Hanlin Sheng$^{1}$, Jun Zhu$^{1}$ and Zefeng Li$^{1}$\\
\normalsize{$^{1}$School of Earth and Space Sciences, University of Science and Technology of China,}\\
\normalsize{Hefei, China.}\\
\\
\normalsize{$^\ast$To whom correspondence should be addressed:}\\
\normalsize{E-mail: xinmwu@ustc.edu.cn}
}
\date{}

%
%

%
%


\maketitle

\begin{abstract}


Training specific deep learning models for particular tasks is common across various domains within seismology. However, this approach encounters two limitations: inadequate labeled data for certain tasks and limited generalization across regions. To address these challenges, we develop SeisCLIP, a seismology foundation model trained through contrastive learning from multi-modal data. It consists of a transformer encoder for extracting crucial features from time-frequency seismic spectrum and an MLP encoder for integrating the phase and source information of the same event. These encoders are jointly pre-trained on a vast dataset and the spectrum encoder is subsequently fine-tuned on smaller datasets for various downstream tasks. Notably, SeisCLIP's performance surpasses that of baseline methods in event classification, localization, and focal mechanism analysis tasks, employing distinct datasets from different regions.  In conclusion, SeisCLIP holds significant potential as a foundational model in the field of seismology, paving the way for innovative directions in foundation-model-based seismology research.

\end{abstract}

\section*{Introduction}

In recent years, the machine-learning or deep-learning based methods has been proved efficiently in almost every subfield of seismology~\citep{beroza2021machine,mousavi2022deep}. These methods have consistently outperformed classical approaches in a wide range of tasks, including but not limited to denoising~\citep{zhu2019seismic, wang2021seismogen, van2021self, yang2022toward, novoselov2022sedenoss}, earthquake detection~\citep{ross2018generalized,yang2021simultaneous,yano2021graph}, phase picking~\citep{ross2018p,mousavi2019cred,pardo2019seismic,wang2019deep,liu2020rapid,bilal2022early,feng2022edgephase,munchmeyer2022picker}, phase association~\citep{ross2019phaselink,mcbrearty2019pairwise,mcbrearty2019earthquake,yu2022fastlink}, localization~\citep{devries2018deep,lomax2019investigation,mousavi2019bayesian,zhang2020locating,van2020automated,zhang2022spatio,mcbrearty2023earthquake}, event classification~\citep{li2018machine, linville2019deep, ku2020attention, kim2021graph,bregman2021array,kong2022combining}, focal mechanism determination~\citep{ross2018p,hara2019p,tian2020comparison,uchide2020focal,kuang2021real,zhu2022deep} and earthquake prediction~\citep{rouet2017machine,wang2021predicting,shokouhi2021deep,johnson2021laboratory,wang2022predicting,borate2023using}. 
Indeed, many existing methods focus on training specific models for individual tasks. To effectively leverage the relationships between related tasks, some researchers have proposed methods to simultaneously address multiple interrelated tasks, such as earthquake detection and picking~\citep{zhu19phasenet,zhu2019deep,mousavi2020earthquake}, earthquake monitoring~\citep{perol2018convolutional, zhu2022end, si2023multi}, localization and magnitude estimation~\citep{munchmeyer2021earthquake}.

Dealing with tasks involving large datasets, training specific models for each task can be effective to a certain extent in obtaining good performance. While there is a vast amount of data in the field of seismology overall, not every specific seismology task has sufficient data samples. The strategy of training a unique model for each task may not work well for those with only limited datasets, such as event classification and focal mechanism determination. To overcome this challenge, transfer learning has been introduced in phase picking~\citep{chai2020using,lapins2021little,zhu2022ustc,niksejel2023obstransformer}, event classification~\citep{titos2019classification,kim2020multifeature} and earthquake peak ground motion prediction~\citep{jozinovic2022transfer}.
Although transfer learning partially mitigates the challenge of data scarcity, its applicability remains constrained to specific sub-fields of seismology.
This issue is not unique in seismology field but is a common challenge across various research domains. To address this problem and enable deep learning methods to better utilize vast amounts of data, an effective way is to train a foundation model through self-supervised learning.

A foundation model is one pre-trained on broad data, typically through supervised learning, and can be adapted (e.g., fine-tuned) to a wide range of downstream tasks~\citep{bommasani2021opportunities} with small datasets.
Meanwhile, self-supervised learning involves pre-training the model on data without the need for human-labeled annotations. 
This approach has been successfully applied in natural language processing (e.g., BERT~\citep{devlin2018bert} and GPT~\citep{brown2020language}) and computer vision (e.g., MAE~\citep{he2022masked}). 
To train a foundation model, there are two main types of self-supervised learning: generative and contrastive. The generative approach involves predicting a part of the data from the rest of the data, while the contrastive approach focuses on learning features based on discriminative signals between different data samples. CLIP~\citep{radford2021learning} is one of the most popular foundation model trained by contrastive approach. 
Trained on multimodal data, it excels in various downstream tasks, such as image generation~\citep{vinker2022clipasso}, segmentation~\citep{xu2022groupvit} and video classification~\citep{luo2022clip4clip}. While most of the research in this area has been focused on image and text data, some researchers have also successfully extended these ideas to audio data, leading to the training of audio foundation models like AST~\citep{gong2021ast}, SSAST~\citep{gong2022ssast}, AudioCLIP~\citep{guzhov2022audioclip} and CAV-MAE~\citep{gong2023contrastive}.

Drawing inspiration from AudioCLIP, AST and recognizing the analogous nature of audio signals and earthquake waveforms as one-dimensional signals, we introduce SeisCLIP, a seismology foundation model pre-trained through contrastive learning on multi-modal data of seismic spectrum and event information (phase and source information). 
We constructed our the foundation model with a two-branch architecture of a Transformer encoder and a simple MLP for the input seismic spectrum and event information, respectively. We built the vast amount of multi-modal training data from the STEAD~\citep{mousavi2019stanford} for pre-training the foundation model. 
In the pre-training process based on contrastive learning, by contrasting the seismic spectrum features extracted by the Transformer encoder and the seismic event information integrated by the MLP encoder, the Transformer encoder will develop the general capability of extracting crucial features that comprehensively represent the seismic data from both local and global perspectives. 
This general feature extraction capability acquired by the Transformer encoder through pre-training means that it can be broadly adapted to various seismic data analysis tasks. 
In all of our evaluation tests on the tasks event classification, location, and focal mechanism analysis associated with different datasets, the pre-trained Transformer encoder consistently outperformed traditional baseline models. This demonstrated the powerful feature extraction capability of the Transformer encoder as well as the versatility and generalizability of this capability. 
To better understand this capability, we visualized and compared the reduced features of the Transformer encoder with the pre-training and other training strategies.
In addition, we discussed the generalization ability of SeisCLIP, considering its potential to be adapted to different datasets and tasks beyond the pre-training domain.

\section*{Methods}

\begin{figure*}[t!]%
\centering
\includegraphics[width=\textwidth]{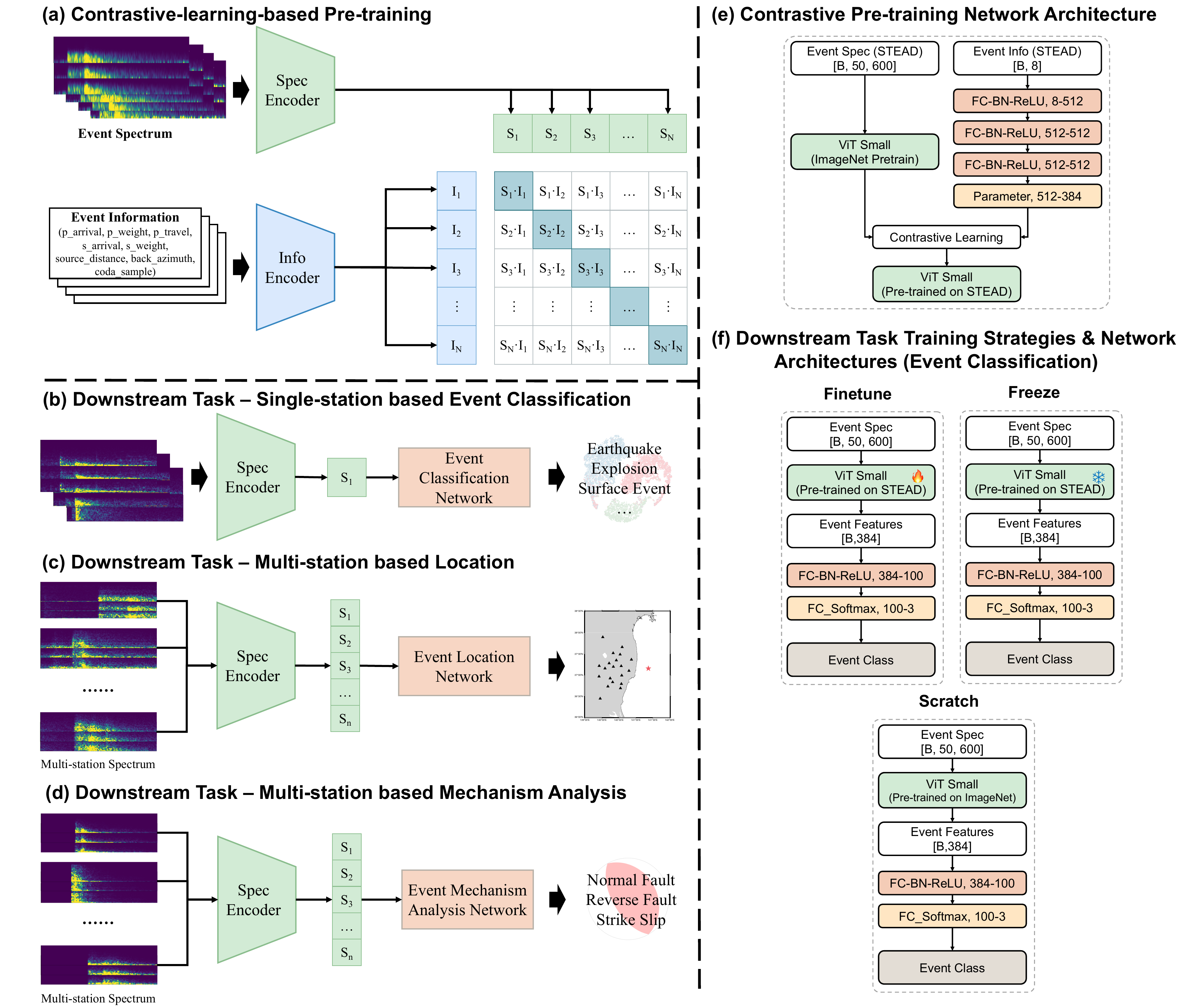}
\caption{Summary of SeisCLIP. (a) SeisCLIP consists of two encoders that are pre-trained jointly by contrastive learning on multi-modal data of seismic spectrum and the corresponding event information (phase and source). (b - d) In downstream tasks, the pre-trained spectrum decoder is employed to generate features, which are then fed to separate sub-networks for event classification, localization, and focal mechanism analysis. (e) Detailed network architectures of the two encoders during the pre-training.
(f) Three different training strategies and the corresponding network architectures of the downstream task of event classification. The detailed network architectures for the other downstream tasks are shown in Figure S2. 
}\label{fig1}
\end{figure*}

\subsection*{Architecture of the Seismology Foundation Model}

Following the principles of CLIP, our goal is to develop a foundational model in seismology using contrastive learning for pre-training. The architecture of foundational models is mostly constructed with Transformer blocks~\citep{vaswani2017attention}, and their training is typically quite time-consuming. To improve computational efficiency, in the case of images, a common approach involves partitioning a single image (often in the size of 256 $\times$ 256) into smaller patches, a series of embedding operations and feeding them into the Transformer network. However, unlike 2D images, seismic waveforms are often 1D long time series, with a waveform often comprising as many as 6000 or even more samples versus an image's 256 $\times$ 256 dimensions. Acknowledging the limitations posed by the length of seismic waveforms, dividing signals into multiple patches can curtail the information contained within each patch. This can also disrupt the continuity of the signal, thus complicating network training. Hence, we refrained from training a waveform-based seismology foundation model. Instead, we adopted a methodology inspired by audio processing techniques~\citep{gong2021ast,gong2022ssast}, to develop a spectrum-based seismology foundation model.

In our foundation model, the network architecture consists of two branches of encoders for input spectrum images and event information (Figure \ref{fig1}a and Figure \ref{fig1}e), respectively. For the input of the spectrum branch, we converted the original seismic waveform data into time-frequency spectrum using the Short-Time Fourier Transform (STFT). This conversion allows the model to work with spectral representations of the seismic signals, enabling it to capture essential frequency-domain features and patterns. For the corresponding spectrum encoder, we utilized the VIT-small ~\citep{dosovitskiy2020image}, pre-trained on the ImageNet dataset, as the backbone. Conversely, the information branch incorporates eight types of event-related phase and source information included in the STEAD dataset: P arrival, P weight, P travel time, S arrival, S weight, source distance, back azimuth, and coda sample. These features were combined and formulated as a 1D vector fed into the information encoder of SeisCLIP. This information encoder is simply a Multi-Layer Perceptron (MLP) consisting of three layers, followed by a single-layer linear projection (Figure \ref{fig1}e). In summary, each training sample consists of two components of a three-channel spectrum data along with its corresponding eight types of event information. These components are separately fed into their respective encoders to generate the spectrum feature and information features. Subsequently, contrastive learning is employed to jointly train the two encorders by comparing the extracted spectrum and information features.

\subsection*{Data Preparation for Pre-training and Validation}

During pre-training, SeisCLIP employs a contrastive training strategy, wherein a spectrum encoder and an information encoder are concurrently trained. To ensure the inclusiveness of the information data, we built the training dataset based on the STEAD dataset, which provides an extensive and diverse set of information related to earthquake events. In total, the STEAD dataset contains around 1,030,000 three-component signals accompanied by complete eight types of event information (phase and source). We employed approximately one million of these instances for training and reserved around 30,000 for validation.

In addition, we evaluated the performance of our pre-trained foundation model by utilizing three distinct downstream tasks: event classification, localization, and focal mechanism analysis. Accordingly, we collected four separate datasets for these tasks. For the event classification task, we employed the PNW dataset, encompassing six event types of earthquake, explosion, surface event, sonic boom, thunder and plane crash. 
However, to ensure balanced sample quantities for each event type, we excluded the small sample categories of plane crash, sonic boom and thunder and randomly selected 20,000 earthquake events from the total dataset. Details about event distribution and sample statistics are provided in Figure S1a and Table S1. In addition, to assess the generalization ability of the pre-trained model, we employed an earthquake-explosion dataset from the southern California seismology network (SCSN), which is a subset of the larger SCSN dataset \citep{zhu2022deep}. Similar to the PNW dataset, the data underwent STFT processing with identical parameters. Specifics on event distribution and sample quantities for this dataset are presented in Figure S1d and Table S2.

For the localization task, we curated data from Japan (Figure S1b), encompassing M~\textgreater2 earthquakes that occurred between January 1, 2011, and December 31, 2011, including the Mw 9.1 Tohoku sequence (Figure S1b). The dataset includes 12,000 events recorded by the 3-component High Sensitivity Seismograph Network (NIED Hi-net). These data were randomly divided into training, validation, and test sets with 10,000, 1,000, and 1,000 samples, respectively.

Furthermore, for the focal mechanism analysis task, we also collected data from Japan (Figure S1c), a region known for a significant number of earthquakes. The dataset comprises some M~\textgreater3 earthquakes occurring between 2011 and 2016, with their focal mechanisms calculated by Japan Meteorological Agency (JMA), totaling more than 2,700 events. The data were randomly divided into training, validation, and test sets (details in Table S1). To maintain consistency with the input format of the pre-trained SeisCLIP model, all waveform data in downstream tasks was cut into 60-second segments with a 100 Hz sample frequency. Spectra were derived through mean-std normalization and transformed into time-frequency representations via STFT.

\subsection*{Pre-training SeisCLIP by Contrastive Learning}

During pre-training, two encoders of SeisCLIP are trained jointly by contrastive learning on multi-modal pairs of seismic spectrum and their corresponding event information (phase and source).
To elaborate further, for a batch of $N$ pairs, the two branches of encoders separately compute the spectrum and information embeddings from their respective inputs. The contrastive-learning-based pre-training is to maximize the cosine similarity of the spectrum and information embeddings for the $N$ real pairs in the batch while minimize the cosine similarity for the $N^2 - N$  incorrect pairings~\citep{radford2021learning}. By optimizing a symmetric cross-entropy loss over these similarity scores, we obtain the pre-trained SeisCLIP model including spectrum encoder and information encoder. 

Throughout the pre-training process, the learning rate was 1e-4, with a batch size of 192. The model was trained for a total of 100 epochs. Due to the use of cross-entropy loss, the model exhibiting the most accurate classification performance on the validation set (55th epoch) was selected as the final pre-trained model. Subsequently, this model is ready to be deployed, utilizing the trained spectrum encoder for various downstream tasks.

\subsection*{Adapting SeisCLIP in Downstream Tasks}

After pre-training, the spectrum encoder of the foundation model serves a threefold purpose in downstream tasks event classification, earthquake localization, and the analysis of seismic source focal mechanisms (Figures \ref{fig1}b-\ref{fig1}d). These tasks correspond respectively to single station classification, multi-station regression, and multi-station classification problems.

In the evaluation of downstream tasks, we employed three distinct strategies: fine-tune, frozen, and scratch. 
The details regarding the training strategy and network architecture for the event classification task 
is shown in Figure 1f. The fine-tune strategy involves activating the pre-trained spectrum encoder (VIT-small) for further training, while the frozen strategy keeps this encoder frozen during the training process. The model from scratch does not utilize the seismology-based pre-trained model.
Moreover, in addition to comparing the pre-trained models using different strategies, we also retrained some baseline models for evaluation. To ensure a fair comparison, we retrained two types of baseline models, one waveform-based and the other spectrum-based, to facilitate comparative analysis for each task.
For the classification task, we utilized the earthquake and explosion classification network~\citep{kong2022combining} for comparison. To accommodate the length of the spectrum data, minor adaptations were made to the network architecture. 

For the localization and focal mechanism analysis tasks, we employed nearly identical baseline networks~\citep{van2020automated}. Architectural divergence exists solely in the activation function of the final layer: sigmoid for location and softmax for focal mechanism analysis. To validate the performance of the foundation model, the sub-networks for the location and focal mechanism tasks closely resemble the baseline models. The key distinction lies in the feature extraction from waveform/spectrum: while the baseline employs multi-layer convolutions, our sub-networks utilize the pre-trained spectrum encoder (Figure S2). To ensure fairness in comparison, all sub-networks and baselines are trained using the same training hyperparameters. Further training details can be found in the Table S3.

\section*{Results}
\subsection*{Event Classification}

\begin{figure*}[t!]%
\centering
\includegraphics[width=0.9\textwidth]{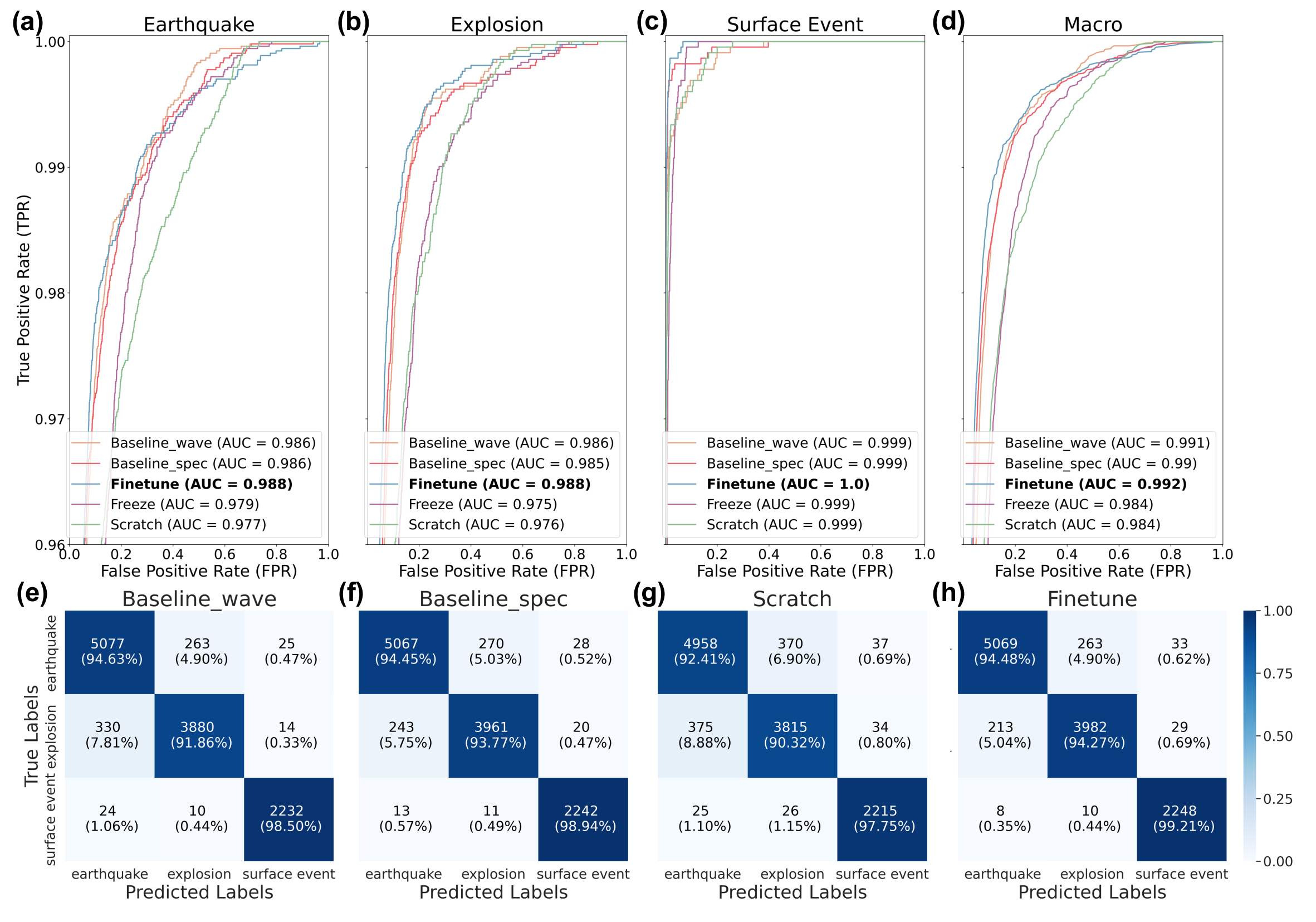}
\caption{Evaluation metrics of different models on the event classification task. (a - c) ROC curves of five different models for the classes of earthquake, explosion, and surface event in the event classification task, respectively. In these five models, the fine-tune, scratch, and frozen models belong to SeisCLIP.  (f) Macro-average represents the arithmetic mean of metrics for each class. The AUC values of each model in each class are indicated in the legend. (e - h) Confusion matrix of four models. Since the frozen model exhibited the poorest performance(the frozen model with the poorest performance is not shown here).
}\label{fig2}
\end{figure*}

For evaluation of the multi-class classification problem, typically one class is designated as the positive sample, while the other types serve as negative samples. Metrics are then calculated individually for each class (details in text S1). Following this approach, we plot the receiver operating characteristic (ROC) curves for each class for the five models (Figures \ref{fig2}a-2d). For all classes, the Area Under Curve (AUC) values of the fine-tune model were consistently better than those of the spectrum-based baseline model. Moreover, across all statistical results, the AUC values of the fine-tune model consistently outperformed the frozen model and the model from scratch.
For a more intuitive analysis, the confusion matrix (Figure \ref{fig2}e-2f) provides a comprehensive view of the performance. The fine-tune model accurately classified a larger number of samples compared to the spectrum-based baseline model. Additionally, due to the pre-training process, the STEAD dataset only contains earthquake events, whereas the downstream task requires classification across three categories. As a result, the scratch model performs better than the frozen model, as it does not suffer from this discrepancy in event types. 

After evaluating the downstream classification tasks in PNW dataset, we further examined the generalization capabilities of different models using an earthquake-explosion dataset from SCSN. With only two classes, AUC curves for binary classification are plotted in Figure S3.
Compared to the PNW dataset, our SeisCLIP demonstrated a more pronounced advantage over baseline models. Notably, the AUC values for both fine-tuned model and model from scratch were higher than those of the baseline models. Additionally, the fine-tuned model attained the highest AUC value among all methods, highlighting SeisCLIP fine-tuned model's exceptional generalization ability.

\subsection*{Event Location}

\begin{figure*}[t!]%
\centering
\includegraphics[width=1.0\textwidth]{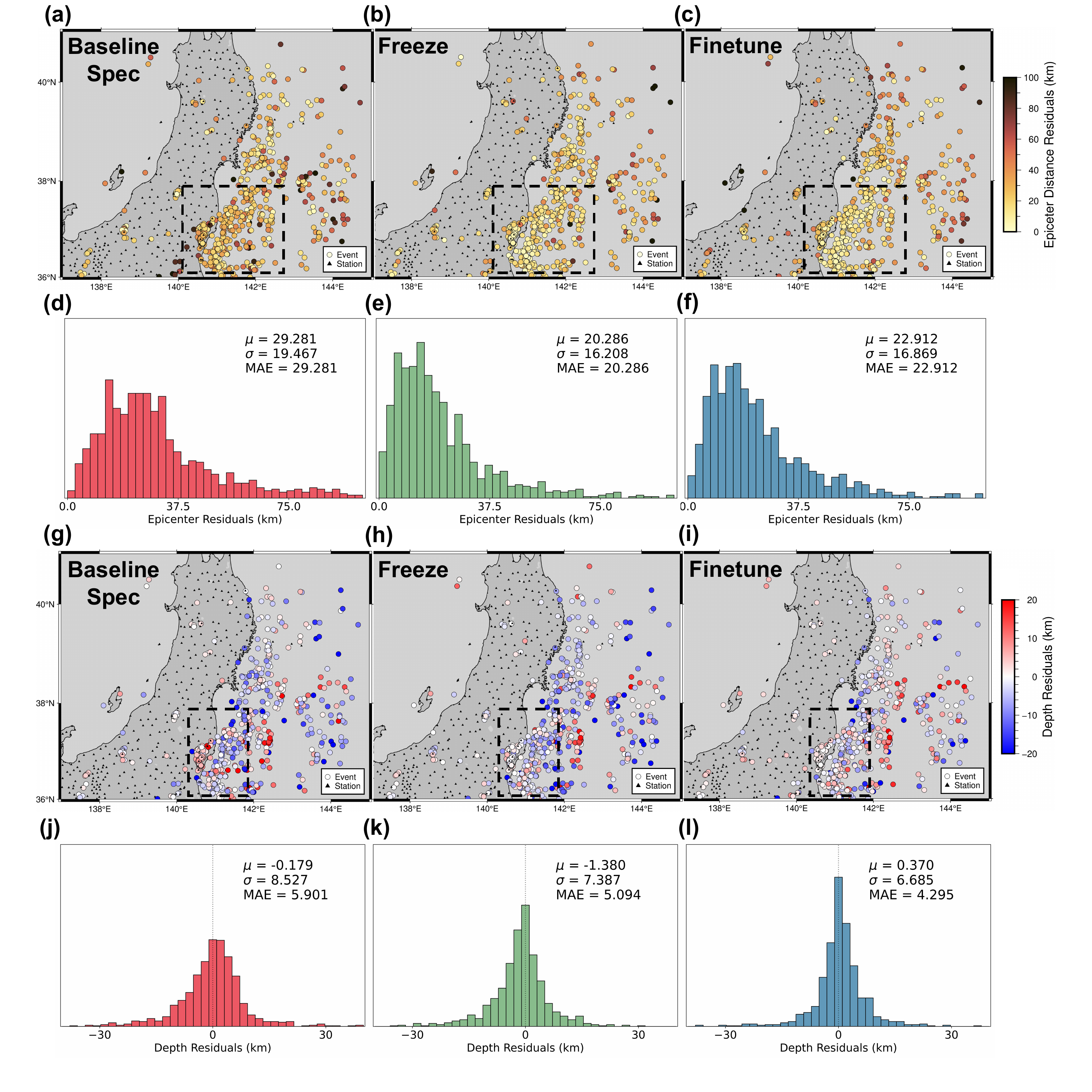}
\caption{Performance of different models on the location task. The subfigures of a - c and g - i display the event locations of test data, where the color of dots represents the location errors. Specifically, the color in a - c corresponds to the epicenter distance residuals, while the color in g - i represents the depth residuals. The subfigures of d - f and j - l illustrate the epicenter/depth residual distributions for each model.
}\label{fig3}
\end{figure*}

Compared to the event classification task, our SeisCLIP method demonstrated a more significant advantage in event localization. This was evident from the residuals in epicenter distance and depth estimation, as depicted in Figure 3. In the first row of Figure 3, we displayed the epicenter distance errors associated with each event, with lighter yellow indicating smaller errors.
The distribution of event points within the highlighted region (black box) of SeisCLIP (both frozen and fine-tune models) appeared shallower compared to the spectrum-based baseline (Figure 3a and 3c). This observation emphasized that the fine-tuned model exhibited lower epicenter localization errors compared to the baseline, highlighting its distinct advantage in accurate epicenter distance estimation.
Moreover, the distribution of epicenter distance errors also supported the above observations. The error distribution of the fine-tuned model is concentrated closer to zero compared to that of the baseline methods, as seen in the second row of Figure 3 and Figure S4, indicating higher overall accuracy. 

In the assessment of depth estimation, we displayed the depth errors associated with each event in the third row of Figure 3, with white denoting smaller errors. In this context, the fine-tuned model notably outperformed the frozen model and baseline model, as evident from the shallower color of event points in the highlighted region (black box) of the fine-tuned model compared to the other methods.
Additionally, these findings are corroborated by the narrowest distribution of residuals and the lowest Mean Absolute Error (MAE) value in the fine-tuned model, as depicted in the fourth row of Figure 3. These results collectively underscored the superior performance of the fine-tuned model in depth estimation.

Beyond epicenter distance and depth residuals, we also analyzed the residual distributions for longitude, latitude and magnitude (depicted in Figure S4). Apart from precise position estimation, SeisCLIP also demonstrated its ability in magnitude estimation. This was evident from the narrower and more concentrated residual distribution in Figure S4, further reinforcing our pre-trained SeisCLIP method's multifaceted advantages in event localization.

\begin{figure*}[t!]%
\centering
\includegraphics[width=\textwidth]{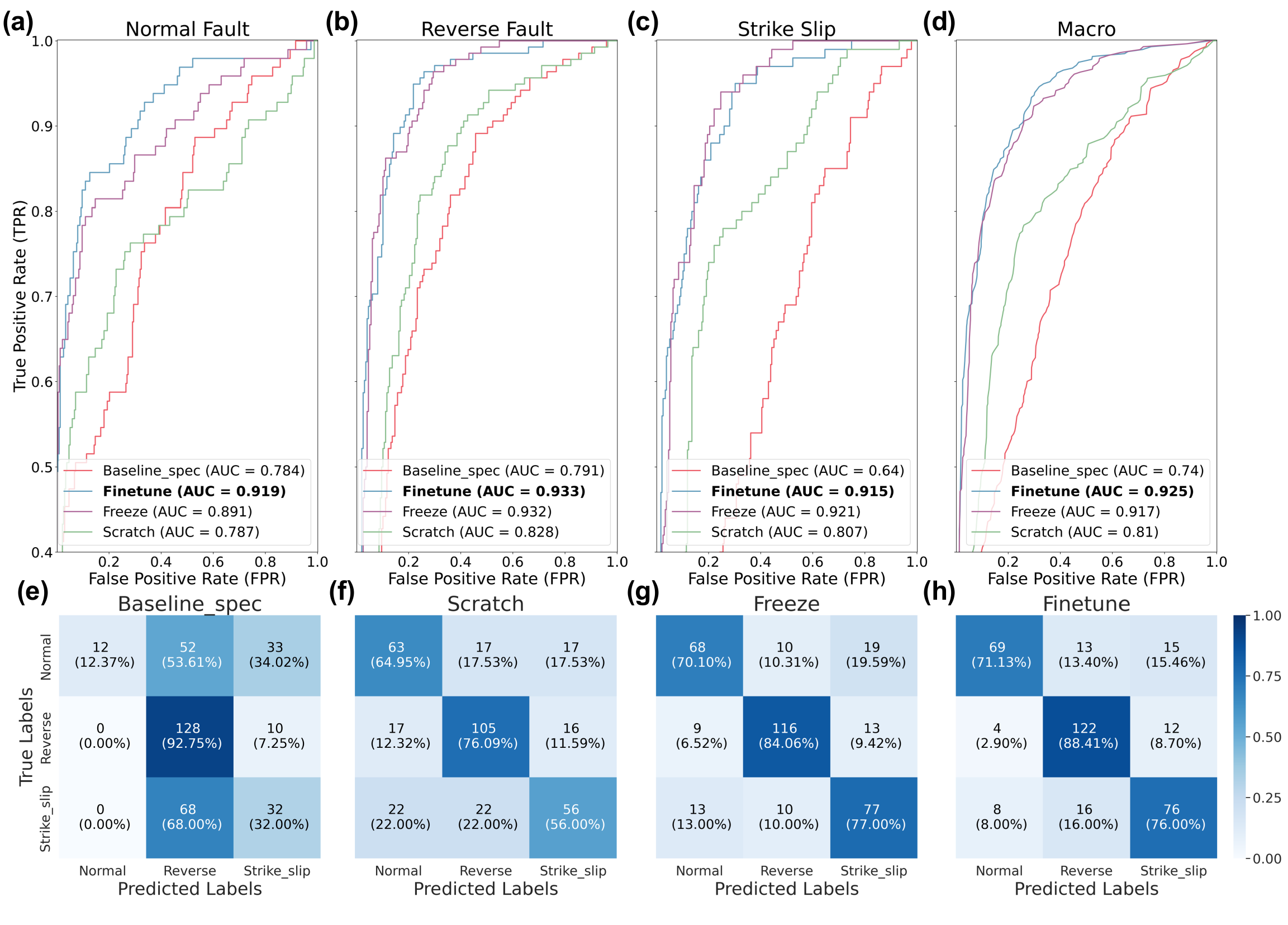}
\caption{Evaluation metrics of four models on the task of focal mechanism analysis: (a - c) the ROC curves for each class of five different models. (d) Macro-average represents the arithmetic mean of metrics for each class. The AUC values of each model in each class are indicated in the legend. (e - h) Confusion matrix of four models.
}\label{fig4}
\end{figure*}

\subsection*{Focal Mechanism Analysis}

Finally, we demonstrated the effectiveness of our foundation model for the focal mechanism analysis task. Given the challenge of determining specific focal mechanisms (strike, dip and rake) with limited training data, we opt to transform this difficult regression problem into a simpler classification problem. 
Here, the task is defined as a multi-station classification problem. It involves inputting multi-station features into the event mechanism analysis network, which then produces an output indicating the type of fault (normal fault, reverse fault, or strike-slip) associated with the event (Figure 1d). For this multi-class classification problem, we applied the same evaluation metrics used in the PNW dataset to assess various methods (Figure 4).

In terms of ROC curves (first row of Figure 4), SeisCLIP the best performance for all the classes of the focal mechanism analysis. Among all methods, the fine-tuned model achieved the highest AUC value, with even the frozen model significantly outperforming the spectrum-based baseline model . The confusion matrix corroborated similar results, with the fine-tuned model exhibiting the highest number of correctly classified samples, while the frozen model performed comparably well to the fine-tuned model. 
It's worth noting that we have treated this task as a multi-station classification problem. Together with the previous single-station classification and multi-station regression tasks of event classification and location, we provided a comprehensive assessment of SeisCLIP's effectiveness.


\section*{Discussion}

\begin{table}[htb!]
\begin{center}
\begin{minipage}{1\textwidth}
\caption{The performance of event classification and location with different spectrum size}\label{tab1}
\begin{tabular*}{\textwidth}{@{\extracolsep{\fill}}cccccccccc@{\extracolsep{\fill}}}
\toprule	
& &  \multicolumn{3}{@{}c@{}}{Event Classificition} & \multicolumn{3}{@{}c@{}}{Location (MAE)} \\ \cline{3-5} \cline{6-8}
Spec Size & Model  & mPre & mRec & mF1 & Epi(km) & Dep (km) & Mag($^\circ$)\\
\midrule
\multirow{4}{*}{50$\times$120}
&Spec-Baseline   & 0.9682 & 0.9691 & 0.9687 & 27.97 & 5.57 & 0.51 \\
& fine-tune (SeisCLIP)   & \textcolor{red}{0.9688} & \textcolor{red}{0.9698} & \textcolor{red}{0.9692} & 25.24 & 
	\textcolor{red}{5.37}  & 0.22\\
& Freeze (SeisCLIP)   &0.9411  & 0.9443  & 0.9426 & \textcolor{red}{24.99} &  
	5.75 & \textcolor{red}{0.20} \\

& Scratch (SeisCLIP)   & 0.9577  & 0.9629  & 0.9602  & 26.96 & 
	6.61 & 0.23\\
	
\midrule
\multirow{4}{*}{50$\times$600}
&Spec-Baseline   & 0.9528 & 0.9546 & 0.9537 & 29.22 & 5.91 & 0.22 \\

& fine-tune (SeisCLIP)   & \textcolor{blue}{0.9557} & \textcolor{blue}{0.9599} & \textcolor{blue}{0.9577} & 22.79 & \textcolor{blue}{4.28}  & \textcolor{blue}{0.19} \\
& Freeze (SeisCLIP)  & 0.9260 & 0.9268 &  0.9263 & \textcolor{blue}{20.27} & 5.10 & 0.22 \\

& Scratch (SeisCLIP)   & 0.9334 & 0.9349 & 0.9342 & 29.42 & 6.66 & 0.32\\

\bottomrule
\end{tabular*}
\footnotetext{Note: The metrics mprecision, mrecall, and mf1 are calculated by averaging the precision, recall, and F1-score values of the three classes. Red color represent the best performance of the spectrum size 50$\times$120. Blue color represent the best performance of the spectrum size 50$\times$600.}
\end{minipage}
\end{center}
\end{table}

By initially pre-training our foundation model and subsequently fine-tuning it cross various downstream tasks, SeisCLIP has demonstrated promising performance in event classification, localization, and focal mechanism analysis. To further elucidate the features captured during the pre-training process, we employed t-SNE~\citep{van2008visualizing} to reduce the features derived from the spectrum decoder into a two-dimensional space. Subsequently, we visualized these reduced features (Figure S5) from the frozen, scratch, and fine-tuned decoders, respectively.

Since we exclusively used earthquake events to pre-train the model, the reduced features from the frozen decoder can only be identified as comcat events (comprising earthquake and explosion events) or exotic events (including surface events). Thus, we can confidently assert that our foundation model learned the intrinsic features from earthquake spectra during the pre-training. 
Moreover, we demonstrated that the fine-tuned model captures more substantial features compared to the model from scratch (as seen in Figure S5b and S5c) after fine-tuning. Specifically, the reduced features of explosions and surface events from the fine-tuned model were distinctly separated from the earthquake features, much more than those of the model from scratch.
This visualization substantiated the efficacy of our pre-trained SeisCLIP model, emphasizing its ability to grasp essential features from earthquake spectra and its enhanced performance after fine-tuning on specific downstream tasks.

Subsequently, we analyzed the influence of different STFT parameters on the model. We set the number of frequency points to 50, spanning a frequency range from 0Hz to 50Hz. Additionally, we employed two different time samples: 120 and 600, corresponding to sampling intervals of 100 and 20, with overlaps of 50 and 10 samples, respectively, at each sampling interval.
Two different foundation models were initially pre-trained using STEAD dataset, each with a different spectrum size. Then, we fine-tuned the 120-sample model and the 600-sample model using the PNW dataset and Japan location dataset respectively, employing their corresponding pre-trained models. 
The ensuing evaluation, utilizing identical metrics (detailed in Table 1), consistently demonstrated our model's superiority over baseline models, regardless of the spectrum size employed. However, it's worth noting that models pre-trained by different spectrum size were suitable for different types of tasks. In the context of classification, the distinctions in event spectra, such as those of earthquakes, explosions, and surface events, are not primarily manifested in temporal sampling. Opting for a smaller spectrum size facilitates easier model learning. On the other hand, for location, which are mainly reliant on arrival time information, employing a larger spectrum size along the temporal dimension allows the network to achieve more precise and accurate outcomes.

The remarkable efficacy of the fine-tuned SeisCLIP model in capturing crucial features across all downstream tasks implies its potential applicability beyond the aforementioned three tasks. Theoretically, our model showcases the capability to address a diverse range of challenges similar to the aforementioned tasks, including but not limited to polarity determination, event detection, and peak ground motion estimation.
However, it is crucial to acknowledge the persistent limitations within our approach. Specifically, our model may encounter challenges in generalization when different-sized spectra were applied to downstream tasks with the pre-trained model. To address this constraint and ensure the adaptability of our model across diverse subfields within seismology, we released open-source pre-trained foundation models of three distinct spectrum sizes, facilitating their customization for various downstream tasks. It is encouraged to explore other strategies to address this limitation and enhance the model's adaptability in more tasks in the future.

\section*{Conclusion}

We have introduced a foundation model in seismology for event classification, location, and focal mechanism analysis. Unlike current deep learning-based methods that require training separate models for each specific task, our proposed foundation model is pre-trained on a large dataset and then fine-tuned on various tasks using smaller datasets. 
In doing so, our foundation model gains a comprehensive understanding of crucial features during the pre-training process but, yielding remarkable performance when adapted to various downstream tasks.

By conducting tests on multiple datasets from the different regions, our method demonstrated its capability to surpass the performance of several baseline models, achieving higher accuracy levels. The versatility showcased by our model across various downstream tasks underscores its potential to address different seismological challenges, setting a blueprint for the development of next-generation seismology deep-learning models.

\section*{Code and Data availability}
 
The open-source foundation model can be found in \url{https://github.com/sixu0/SeisCLIP/}. The STEAD and PNW dataset can be downloaded from \url{https://github.com/smousavi05/STEAD} and \url{https://github.com/niyiyu/PNW-ML}. The event waveform and focal mechanism for Japan can be downloaded from HiNet (\url{https://www.hinet.bosai.go.jp/?LANG=en}) and the Japan Meteorological Agency (\url{https://www.data.jma.go.jp/svd/eqev/data/bulletin/index}).


%
%
\bibliography{si23SeisCLIP.bib}

\begin{thebibliography}{76}
\providecommand{\natexlab}[1]{#1}
\providecommand{\url}[1]{\texttt{#1}}
\expandafter\ifx\csname urlstyle\endcsname\relax
  \providecommand{\doi}[1]{doi: #1}\else
  \providecommand{\doi}{doi: \begingroup \urlstyle{rm}\Url}\fi

\bibitem[Beroza et~al.(2021)Beroza, Segou, and
  Mostafa~Mousavi]{beroza2021machine}
Gregory~C Beroza, Margarita Segou, and S~Mostafa~Mousavi.
\newblock Machine learning and earthquake forecasting—next steps.
\newblock \emph{Nature communications}, 12\penalty0 (1):\penalty0 4761, 2021.

\bibitem[Bilal et~al.(2022)Bilal, Ji, Wang, Akhter, and Yaqub]{bilal2022early}
Muhammad~Atif Bilal, Yanju Ji, Yongzhi Wang, Muhammad~Pervez Akhter, and
  Muhammad Yaqub.
\newblock Early earthquake detection using batch normalization graph
  convolutional neural network (bngcnn).
\newblock \emph{Applied Sciences}, 12\penalty0 (15):\penalty0 7548, 2022.

\bibitem[Bommasani et~al.(2021)Bommasani, Hudson, Adeli, Altman, Arora, von
  Arx, Bernstein, Bohg, Bosselut, Brunskill,
  et~al.]{bommasani2021opportunities}
Rishi Bommasani, Drew~A Hudson, Ehsan Adeli, Russ Altman, Simran Arora, Sydney
  von Arx, Michael~S Bernstein, Jeannette Bohg, Antoine Bosselut, Emma
  Brunskill, et~al.
\newblock On the opportunities and risks of foundation models.
\newblock \emph{arXiv preprint arXiv:2108.07258}, 2021.

\bibitem[Borate et~al.(2023)Borate, Rivi{\`e}re, Marone, Mali, Kifer, and
  Shokouhi]{borate2023using}
Prabhav Borate, Jacques Rivi{\`e}re, Chris Marone, Ankur Mali, Daniel Kifer,
  and Parisa Shokouhi.
\newblock Using a physics-informed neural network and fault zone acoustic
  monitoring to predict lab earthquakes.
\newblock \emph{Nature Communications}, 14\penalty0 (1):\penalty0 3693, 2023.

\bibitem[Bregman et~al.(2021)Bregman, Lindenbaum, and Rabin]{bregman2021array}
Y~Bregman, O~Lindenbaum, and N~Rabin.
\newblock Array based earthquakes-explosion discrimination using diffusion
  maps.
\newblock \emph{Pure and Applied Geophysics}, 178:\penalty0 2403--2418, 2021.

\bibitem[Brown et~al.(2020)Brown, Mann, Ryder, Subbiah, Kaplan, Dhariwal,
  Neelakantan, Shyam, Sastry, Askell, et~al.]{brown2020language}
Tom Brown, Benjamin Mann, Nick Ryder, Melanie Subbiah, Jared~D Kaplan, Prafulla
  Dhariwal, Arvind Neelakantan, Pranav Shyam, Girish Sastry, Amanda Askell,
  et~al.
\newblock Language models are few-shot learners.
\newblock \emph{Advances in neural information processing systems},
  33:\penalty0 1877--1901, 2020.

\bibitem[Chai et~al.(2020)Chai, Maceira, Santos-Villalobos, Venkatakrishnan,
  Schoenball, Zhu, Beroza, Thurber, and Team]{chai2020using}
Chengping Chai, Monica Maceira, Hector~J Santos-Villalobos, Singanallur~V
  Venkatakrishnan, Martin Schoenball, Weiqiang Zhu, Gregory~C Beroza, Clifford
  Thurber, and EGS~Collab Team.
\newblock Using a deep neural network and transfer learning to bridge scales
  for seismic phase picking.
\newblock \emph{Geophysical Research Letters}, 47\penalty0 (16):\penalty0
  e2020GL088651, 2020.

\bibitem[Devlin et~al.(2018)Devlin, Chang, Lee, and Toutanova]{devlin2018bert}
Jacob Devlin, Ming-Wei Chang, Kenton Lee, and Kristina Toutanova.
\newblock Bert: Pre-training of deep bidirectional transformers for language
  understanding.
\newblock \emph{arXiv preprint arXiv:1810.04805}, 2018.

\bibitem[DeVries et~al.(2018)DeVries, Vi{\'e}gas, Wattenberg, and
  Meade]{devries2018deep}
Phoebe~MR DeVries, Fernanda Vi{\'e}gas, Martin Wattenberg, and Brendan~J Meade.
\newblock Deep learning of aftershock patterns following large earthquakes.
\newblock \emph{Nature}, 560\penalty0 (7720):\penalty0 632--634, 2018.

\bibitem[Dosovitskiy et~al.(2020)Dosovitskiy, Beyer, Kolesnikov, Weissenborn,
  Zhai, Unterthiner, Dehghani, Minderer, Heigold, Gelly,
  et~al.]{dosovitskiy2020image}
Alexey Dosovitskiy, Lucas Beyer, Alexander Kolesnikov, Dirk Weissenborn,
  Xiaohua Zhai, Thomas Unterthiner, Mostafa Dehghani, Matthias Minderer, Georg
  Heigold, Sylvain Gelly, et~al.
\newblock An image is worth 16x16 words: Transformers for image recognition at
  scale.
\newblock \emph{arXiv preprint arXiv:2010.11929}, 2020.

\bibitem[Feng et~al.(2022)Feng, Mohanna, and Meng]{feng2022edgephase}
Tian Feng, Saeed Mohanna, and Lingsen Meng.
\newblock Edgephase: A deep learning model for multi-station seismic phase
  picking.
\newblock \emph{Geochemistry, Geophysics, Geosystems}, 23\penalty0
  (11):\penalty0 e2022GC010453, 2022.

\bibitem[Gong et~al.(2021)Gong, Chung, and Glass]{gong2021ast}
Yuan Gong, Yu-An Chung, and James Glass.
\newblock Ast: Audio spectrogram transformer.
\newblock \emph{arXiv preprint arXiv:2104.01778}, 2021.

\bibitem[Gong et~al.(2022)Gong, Lai, Chung, and Glass]{gong2022ssast}
Yuan Gong, Cheng-I Lai, Yu-An Chung, and James Glass.
\newblock Ssast: Self-supervised audio spectrogram transformer.
\newblock In \emph{Proceedings of the AAAI Conference on Artificial
  Intelligence}, volume~36, pages 10699--10709, 2022.

\bibitem[Gong et~al.(2023)Gong, Rouditchenko, Liu, Harwath, Karlinsky, Kuehne,
  and Glass]{gong2023contrastive}
Yuan Gong, Andrew Rouditchenko, Alexander~H. Liu, David Harwath, Leonid
  Karlinsky, Hilde Kuehne, and James~R. Glass.
\newblock Contrastive audio-visual masked autoencoder.
\newblock In \emph{The Eleventh International Conference on Learning
  Representations}, 2023.

\bibitem[Guzhov et~al.(2022)Guzhov, Raue, Hees, and
  Dengel]{guzhov2022audioclip}
Andrey Guzhov, Federico Raue, J{\"o}rn Hees, and Andreas Dengel.
\newblock Audioclip: Extending clip to image, text and audio.
\newblock In \emph{ICASSP 2022-2022 IEEE International Conference on Acoustics,
  Speech and Signal Processing (ICASSP)}, pages 976--980. IEEE, 2022.

\bibitem[Hara et~al.(2019)Hara, Fukahata, and Iio]{hara2019p}
Shota Hara, Yukitoshi Fukahata, and Yoshihisa Iio.
\newblock P-wave first-motion polarity determination of waveform data in
  western japan using deep learning.
\newblock \emph{Earth, Planets and Space}, 71\penalty0 (1):\penalty0 1--11,
  2019.

\bibitem[He et~al.(2022)He, Chen, Xie, Li, Doll{\'a}r, and
  Girshick]{he2022masked}
Kaiming He, Xinlei Chen, Saining Xie, Yanghao Li, Piotr Doll{\'a}r, and Ross
  Girshick.
\newblock Masked autoencoders are scalable vision learners.
\newblock In \emph{Proceedings of the IEEE/CVF conference on computer vision
  and pattern recognition}, pages 16000--16009, 2022.

\bibitem[Johnson et~al.(2021)Johnson, Rouet-Leduc, Pyrak-Nolte, Beroza, Marone,
  Hulbert, Howard, Singer, Gordeev, Karaflos, et~al.]{johnson2021laboratory}
Paul~A Johnson, Bertrand Rouet-Leduc, Laura~J Pyrak-Nolte, Gregory~C Beroza,
  Chris~J Marone, Claudia Hulbert, Addison Howard, Philipp Singer, Dmitry
  Gordeev, Dimosthenis Karaflos, et~al.
\newblock Laboratory earthquake forecasting: A machine learning competition.
\newblock \emph{Proceedings of the national academy of sciences}, 118\penalty0
  (5):\penalty0 e2011362118, 2021.

\bibitem[Jozinovi{\'c} et~al.(2022)Jozinovi{\'c}, Lomax, {\v{S}}tajduhar, and
  Michelini]{jozinovic2022transfer}
Dario Jozinovi{\'c}, Anthony Lomax, Ivan {\v{S}}tajduhar, and Alberto
  Michelini.
\newblock Transfer learning: Improving neural network based prediction of
  earthquake ground shaking for an area with insufficient training data.
\newblock \emph{Geophysical Journal International}, 229\penalty0 (1):\penalty0
  704--718, 2022.

\bibitem[Kim et~al.(2020)Kim, Ku, and Ko]{kim2020multifeature}
Gwantae Kim, Bonhwa Ku, and Hanseok Ko.
\newblock Multifeature fusion-based earthquake event classification using
  transfer learning.
\newblock \emph{IEEE Geoscience and Remote Sensing Letters}, 18\penalty0
  (6):\penalty0 974--978, 2020.

\bibitem[Kim et~al.(2021)Kim, Ku, Ahn, and Ko]{kim2021graph}
Gwantae Kim, Bonhwa Ku, Jae-Kwang Ahn, and Hanseok Ko.
\newblock Graph convolution networks for seismic events classification using
  raw waveform data from multiple stations.
\newblock \emph{IEEE Geoscience and Remote Sensing Letters}, 19:\penalty0 1--5,
  2021.

\bibitem[Kong et~al.(2022)Kong, Wang, Walter, Pyle, Koper, and
  Schmandt]{kong2022combining}
Qingkai Kong, Ruijia Wang, William~R Walter, Moira Pyle, Keith Koper, and
  Brandon Schmandt.
\newblock Combining deep learning with physics based features in
  explosion-earthquake discrimination.
\newblock \emph{Geophysical Research Letters}, 49\penalty0 (13):\penalty0
  e2022GL098645, 2022.

\bibitem[Ku et~al.(2020)Ku, Kim, Ahn, Lee, and Ko]{ku2020attention}
Bonhwa Ku, Gwantae Kim, Jae-Kwang Ahn, Jimin Lee, and Hanseok Ko.
\newblock Attention-based convolutional neural network for earthquake event
  classification.
\newblock \emph{IEEE Geoscience and Remote Sensing Letters}, 18\penalty0
  (12):\penalty0 2057--2061, 2020.

\bibitem[Kuang et~al.(2021)Kuang, Yuan, and Zhang]{kuang2021real}
Wenhuan Kuang, Congcong Yuan, and Jie Zhang.
\newblock Real-time determination of earthquake focal mechanism via deep
  learning.
\newblock \emph{Nature communications}, 12\penalty0 (1):\penalty0 1--8, 2021.

\bibitem[Lapins et~al.(2021)Lapins, Goitom, Kendall, Werner, Cashman, and
  Hammond]{lapins2021little}
Sacha Lapins, Berhe Goitom, J-Michael Kendall, Maximilian~J Werner, Katharine~V
  Cashman, and James~OS Hammond.
\newblock A little data goes a long way: Automating seismic phase arrival
  picking at nabro volcano with transfer learning.
\newblock \emph{Journal of Geophysical Research: Solid Earth}, 126\penalty0
  (7):\penalty0 e2021JB021910, 2021.

\bibitem[Li et~al.(2018)Li, Meier, Hauksson, Zhan, and Andrews]{li2018machine}
Zefeng Li, Men-Andrin Meier, Egill Hauksson, Zhongwen Zhan, and Jennifer
  Andrews.
\newblock Machine learning seismic wave discrimination: Application to
  earthquake early warning.
\newblock \emph{Geophysical Research Letters}, 45\penalty0 (10):\penalty0
  4773--4779, 2018.

\bibitem[Linville et~al.(2019)Linville, Pankow, and Draelos]{linville2019deep}
Lisa Linville, Kristine Pankow, and Timothy Draelos.
\newblock Deep learning models augment analyst decisions for event
  discrimination.
\newblock \emph{Geophysical Research Letters}, 46\penalty0 (7):\penalty0
  3643--3651, 2019.

\bibitem[Liu et~al.(2020)Liu, Zhang, Zhu, Ellsworth, and Li]{liu2020rapid}
Min Liu, Miao Zhang, Weiqiang Zhu, William~L Ellsworth, and Hongyi Li.
\newblock Rapid characterization of the july 2019 ridgecrest, california,
  earthquake sequence from raw seismic data using machine-learning phase
  picker.
\newblock \emph{Geophysical Research Letters}, 47\penalty0 (4):\penalty0
  e2019GL086189, 2020.

\bibitem[Lomax et~al.(2019)Lomax, Michelini, and
  Jozinovi{\'c}]{lomax2019investigation}
Anthony Lomax, Alberto Michelini, and Dario Jozinovi{\'c}.
\newblock An investigation of rapid earthquake characterization using
  single-station waveforms and a convolutional neural network.
\newblock \emph{Seismological Research Letters}, 90\penalty0 (2A):\penalty0
  517--529, 2019.

\bibitem[Luo et~al.(2022)Luo, Ji, Zhong, Chen, Lei, Duan, and
  Li]{luo2022clip4clip}
Huaishao Luo, Lei Ji, Ming Zhong, Yang Chen, Wen Lei, Nan Duan, and Tianrui Li.
\newblock Clip4clip: An empirical study of clip for end to end video clip
  retrieval and captioning.
\newblock \emph{Neurocomputing}, 508:\penalty0 293--304, 2022.

\bibitem[McBrearty and Beroza(2023)]{mcbrearty2023earthquake}
Ian~W McBrearty and Gregory~C Beroza.
\newblock Earthquake phase association with graph neural networks.
\newblock \emph{Bulletin of the Seismological Society of America}, 113\penalty0
  (2):\penalty0 524--547, 2023.

\bibitem[McBrearty et~al.(2019{\natexlab{a}})McBrearty, Delorey, and
  Johnson]{mcbrearty2019pairwise}
Ian~W McBrearty, Andrew~A Delorey, and Paul~A Johnson.
\newblock Pairwise association of seismic arrivals with convolutional neural
  networks.
\newblock \emph{Seismological Research Letters}, 90\penalty0 (2A):\penalty0
  503--509, 2019{\natexlab{a}}.

\bibitem[McBrearty et~al.(2019{\natexlab{b}})McBrearty, Gomberg, Delorey, and
  Johnson]{mcbrearty2019earthquake}
Ian~W McBrearty, Joan Gomberg, Andrew~A Delorey, and Paul~A Johnson.
\newblock Earthquake arrival association with backprojection and graph
  theoryearthquake arrival association with backprojection and graph theory.
\newblock \emph{Bulletin of the Seismological Society of America}, 109\penalty0
  (6):\penalty0 2510--2531, 2019{\natexlab{b}}.

\bibitem[Mousavi and Beroza(2019)]{mousavi2019bayesian}
S~Mostafa Mousavi and Gregory~C Beroza.
\newblock Bayesian-deep-learning estimation of earthquake location from
  single-station observations.
\newblock \emph{arXiv preprint arXiv:1912.01144}, 2019.

\bibitem[Mousavi and Beroza(2022)]{mousavi2022deep}
S~Mostafa Mousavi and Gregory~C Beroza.
\newblock Deep-learning seismology.
\newblock \emph{Science}, 377\penalty0 (6607):\penalty0 eabm4470, 2022.

\bibitem[Mousavi et~al.(2019{\natexlab{a}})Mousavi, Sheng, Zhu, and
  Beroza]{mousavi2019stanford}
S~Mostafa Mousavi, Yixiao Sheng, Weiqiang Zhu, and Gregory~C Beroza.
\newblock Stanford earthquake dataset (stead): A global data set of seismic
  signals for ai.
\newblock \emph{IEEE Access}, 7:\penalty0 179464--179476, 2019{\natexlab{a}}.

\bibitem[Mousavi et~al.(2019{\natexlab{b}})Mousavi, Zhu, Sheng, and
  Beroza]{mousavi2019cred}
S~Mostafa Mousavi, Weiqiang Zhu, Yixiao Sheng, and Gregory~C Beroza.
\newblock Cred: A deep residual network of convolutional and recurrent units
  for earthquake signal detection.
\newblock \emph{Scientific reports}, 9\penalty0 (1):\penalty0 1--14,
  2019{\natexlab{b}}.

\bibitem[Mousavi et~al.(2020)Mousavi, Ellsworth, Zhu, Chuang, and
  Beroza]{mousavi2020earthquake}
S~Mostafa Mousavi, William~L Ellsworth, Weiqiang Zhu, Lindsay~Y Chuang, and
  Gregory~C Beroza.
\newblock Earthquake transformer—an attentive deep-learning model for
  simultaneous earthquake detection and phase picking.
\newblock \emph{Nature communications}, 11\penalty0 (1):\penalty0 1--12, 2020.

\bibitem[M{\"u}nchmeyer et~al.(2021)M{\"u}nchmeyer, Bindi, Leser, and
  Tilmann]{munchmeyer2021earthquake}
Jannes M{\"u}nchmeyer, Dino Bindi, Ulf Leser, and Frederik Tilmann.
\newblock Earthquake magnitude and location estimation from real time seismic
  waveforms with a transformer network.
\newblock \emph{Geophysical Journal International}, 226\penalty0 (2):\penalty0
  1086--1104, 2021.

\bibitem[M{\"u}nchmeyer et~al.(2022)M{\"u}nchmeyer, Woollam, Rietbrock,
  Tilmann, Lange, Bornstein, Diehl, Giunchi, Haslinger, Jozinovi{\'c},
  et~al.]{munchmeyer2022picker}
Jannes M{\"u}nchmeyer, Jack Woollam, Andreas Rietbrock, Frederik Tilmann,
  Dietrich Lange, Thomas Bornstein, Tobias Diehl, Carlo Giunchi, Florian
  Haslinger, Dario Jozinovi{\'c}, et~al.
\newblock Which picker fits my data? a quantitative evaluation of deep learning
  based seismic pickers.
\newblock \emph{Journal of Geophysical Research: Solid Earth}, 127\penalty0
  (1):\penalty0 e2021JB023499, 2022.

\bibitem[Niksejel and Zhang(2023)]{niksejel2023obstransformer}
Alireza Niksejel and Miao Zhang.
\newblock Obstransformer: A deep-learning seismic phase picker for obs data
  using automated labelling and transfer learning.
\newblock \emph{arXiv preprint arXiv:2306.04753}, 2023.

\bibitem[Novoselov et~al.(2022)Novoselov, Balazs, and
  Bokelmann]{novoselov2022sedenoss}
Artemii Novoselov, Peter Balazs, and G{\"o}tz Bokelmann.
\newblock Sedenoss: Separating and denoising seismic signals with dual-path
  recurrent neural network architecture.
\newblock \emph{Journal of Geophysical Research: Solid Earth}, 127\penalty0
  (3):\penalty0 e2021JB023183, 2022.

\bibitem[Pardo et~al.(2019)Pardo, Garfias, and Malpica]{pardo2019seismic}
Esteban Pardo, Carmen Garfias, and Norberto Malpica.
\newblock Seismic phase picking using convolutional networks.
\newblock \emph{IEEE Transactions on Geoscience and Remote Sensing},
  57\penalty0 (9):\penalty0 7086--7092, 2019.

\bibitem[Perol et~al.(2018)Perol, Gharbi, and Denolle]{perol2018convolutional}
Thibaut Perol, Micha{\"e}l Gharbi, and Marine Denolle.
\newblock Convolutional neural network for earthquake detection and location.
\newblock \emph{Science Advances}, 4\penalty0 (2):\penalty0 e1700578, 2018.

\bibitem[Radford et~al.(2021)Radford, Kim, Hallacy, Ramesh, Goh, Agarwal,
  Sastry, Askell, Mishkin, Clark, et~al.]{radford2021learning}
Alec Radford, Jong~Wook Kim, Chris Hallacy, Aditya Ramesh, Gabriel Goh,
  Sandhini Agarwal, Girish Sastry, Amanda Askell, Pamela Mishkin, Jack Clark,
  et~al.
\newblock Learning transferable visual models from natural language
  supervision.
\newblock In \emph{International conference on machine learning}, pages
  8748--8763. PMLR, 2021.

\bibitem[Ross et~al.(2018{\natexlab{a}})Ross, Meier, and Hauksson]{ross2018p}
Zachary~E Ross, Men-Andrin Meier, and Egill Hauksson.
\newblock P wave arrival picking and first-motion polarity determination with
  deep learning.
\newblock \emph{Journal of Geophysical Research: Solid Earth}, 123\penalty0
  (6):\penalty0 5120--5129, 2018{\natexlab{a}}.

\bibitem[Ross et~al.(2018{\natexlab{b}})Ross, Meier, Hauksson, and
  Heaton]{ross2018generalized}
Zachary~E Ross, Men-Andrin Meier, Egill Hauksson, and Thomas~H Heaton.
\newblock Generalized seismic phase detection with deep learningshort note.
\newblock \emph{Bulletin of the Seismological Society of America}, 108\penalty0
  (5A):\penalty0 2894--2901, 2018{\natexlab{b}}.

\bibitem[Ross et~al.(2019)Ross, Yue, Meier, Hauksson, and
  Heaton]{ross2019phaselink}
Zachary~E Ross, Yisong Yue, Men-Andrin Meier, Egill Hauksson, and Thomas~H
  Heaton.
\newblock Phaselink: A deep learning approach to seismic phase association.
\newblock \emph{Journal of Geophysical Research: Solid Earth}, 124\penalty0
  (1):\penalty0 856--869, 2019.

\bibitem[Rouet-Leduc et~al.(2017)Rouet-Leduc, Hulbert, Lubbers, Barros,
  Humphreys, and Johnson]{rouet2017machine}
Bertrand Rouet-Leduc, Claudia Hulbert, Nicholas Lubbers, Kipton Barros, Colin~J
  Humphreys, and Paul~A Johnson.
\newblock Machine learning predicts laboratory earthquakes.
\newblock \emph{Geophysical Research Letters}, 44\penalty0 (18):\penalty0
  9276--9282, 2017.

\bibitem[Shokouhi et~al.(2021)Shokouhi, Girkar, Rivi{\`e}re, Shreedharan,
  Marone, Giles, and Kifer]{shokouhi2021deep}
Parisa Shokouhi, Vrushali Girkar, Jacques Rivi{\`e}re, Srisharan Shreedharan,
  Chris Marone, C~Lee Giles, and Daniel Kifer.
\newblock Deep learning can predict laboratory quakes from active source
  seismic data.
\newblock \emph{Geophysical Research Letters}, 48\penalty0 (12):\penalty0
  e2021GL093187, 2021.

\bibitem[Si et~al.(2023)Si, Wu, Li, Wang, and Zhu]{si2023multi}
Xu~Si, Xinming Wu, Zefeng Li, Shenghou Wang, and Jun Zhu.
\newblock Multi-task multi-station earthquake monitoring: An all-in-one seismic
  phase picking, location, and association network (plan).
\newblock \emph{arXiv preprint arXiv:2306.13918}, 2023.

\bibitem[Tian et~al.(2020)Tian, Zhang, Zhang, Zhang, Zhang, Wang, and
  Guo]{tian2020comparison}
Xiao Tian, Wei Zhang, Xiong Zhang, Jie Zhang, Qingshan Zhang, Xiangteng Wang,
  and Quanshi Guo.
\newblock Comparison of single-trace and multiple-trace polarity determination
  for surface microseismic data using deep learning.
\newblock \emph{Seismological Research Letters}, 91\penalty0 (3):\penalty0
  1794--1803, 2020.

\bibitem[Titos et~al.(2019)Titos, Bueno, Garc{\'\i}a, Ben{\'\i}tez, and
  Segura]{titos2019classification}
Manuel Titos, Angel Bueno, Luz Garc{\'\i}a, Carmen Ben{\'\i}tez, and Jos{\'e}~C
  Segura.
\newblock Classification of isolated volcano-seismic events based on inductive
  transfer learning.
\newblock \emph{IEEE Geoscience and Remote Sensing Letters}, 17\penalty0
  (5):\penalty0 869--873, 2019.

\bibitem[Uchide(2020)]{uchide2020focal}
Takahiko Uchide.
\newblock Focal mechanisms of small earthquakes beneath the japanese islands
  based on first-motion polarities picked using deep learning.
\newblock \emph{Geophysical Journal International}, 223\penalty0 (3):\penalty0
  1658--1671, 2020.

\bibitem[van~den Ende et~al.(2021)van~den Ende, Lior, Ampuero, Sladen, Ferrari,
  and Richard]{van2021self}
Martijn van~den Ende, Itzhak Lior, Jean-Paul Ampuero, Anthony Sladen, Andr{\'e}
  Ferrari, and C{\'e}dric Richard.
\newblock A self-supervised deep learning approach for blind denoising and
  waveform coherence enhancement in distributed acoustic sensing data.
\newblock \emph{IEEE Transactions on Neural Networks and Learning Systems},
  2021.

\bibitem[van~den Ende and Ampuero(2020)]{van2020automated}
Martijn~PA van~den Ende and J-P Ampuero.
\newblock Automated seismic source characterization using deep graph neural
  networks.
\newblock \emph{Geophysical Research Letters}, 47\penalty0 (17):\penalty0
  e2020GL088690, 2020.

\bibitem[Van~der Maaten and Hinton(2008)]{van2008visualizing}
Laurens Van~der Maaten and Geoffrey Hinton.
\newblock Visualizing data using t-sne.
\newblock \emph{Journal of machine learning research}, 9\penalty0 (11), 2008.

\bibitem[Vaswani et~al.(2017)Vaswani, Shazeer, Parmar, Uszkoreit, Jones, Gomez,
  Kaiser, and Polosukhin]{vaswani2017attention}
Ashish Vaswani, Noam Shazeer, Niki Parmar, Jakob Uszkoreit, Llion Jones,
  Aidan~N Gomez, {\L}ukasz Kaiser, and Illia Polosukhin.
\newblock Attention is all you need.
\newblock \emph{Advances in neural information processing systems}, 30, 2017.

\bibitem[Vinker et~al.(2022)Vinker, Pajouheshgar, Bo, Bachmann, Bermano,
  Cohen-Or, Zamir, and Shamir]{vinker2022clipasso}
Yael Vinker, Ehsan Pajouheshgar, Jessica~Y Bo, Roman~Christian Bachmann,
  Amit~Haim Bermano, Daniel Cohen-Or, Amir Zamir, and Ariel Shamir.
\newblock Clipasso: Semantically-aware object sketching.
\newblock \emph{ACM Transactions on Graphics (TOG)}, 41\penalty0 (4):\penalty0
  1--11, 2022.

\bibitem[Wang et~al.(2019)Wang, Xiao, Liu, Zhao, and Yao]{wang2019deep}
Jian Wang, Zhuowei Xiao, Chang Liu, Dapeng Zhao, and Zhenxing Yao.
\newblock Deep learning for picking seismic arrival times.
\newblock \emph{Journal of Geophysical Research: Solid Earth}, 124\penalty0
  (7):\penalty0 6612--6624, 2019.

\bibitem[Wang et~al.(2021{\natexlab{a}})Wang, Johnson, Bennett, and
  Johnson]{wang2021predicting}
Kun Wang, Christopher~W Johnson, Kane~C Bennett, and Paul~A Johnson.
\newblock Predicting fault slip via transfer learning.
\newblock \emph{Nature Communications}, 12\penalty0 (1):\penalty0 7319,
  2021{\natexlab{a}}.

\bibitem[Wang et~al.(2022)Wang, Johnson, Bennett, and
  Johnson]{wang2022predicting}
Kun Wang, Christopher~W Johnson, Kane~C Bennett, and Paul~A Johnson.
\newblock Predicting future laboratory fault friction through deep learning
  transformer models.
\newblock \emph{Geophysical Research Letters}, 49\penalty0 (19):\penalty0
  e2022GL098233, 2022.

\bibitem[Wang et~al.(2021{\natexlab{b}})Wang, Trugman, and
  Lin]{wang2021seismogen}
Tiantong Wang, Daniel Trugman, and Youzuo Lin.
\newblock Seismogen: Seismic waveform synthesis using gan with application to
  seismic data augmentation.
\newblock \emph{Journal of Geophysical Research: Solid Earth}, 126\penalty0
  (4):\penalty0 e2020JB020077, 2021{\natexlab{b}}.

\bibitem[Xu et~al.(2022)Xu, De~Mello, Liu, Byeon, Breuel, Kautz, and
  Wang]{xu2022groupvit}
Jiarui Xu, Shalini De~Mello, Sifei Liu, Wonmin Byeon, Thomas Breuel, Jan Kautz,
  and Xiaolong Wang.
\newblock Groupvit: Semantic segmentation emerges from text supervision.
\newblock In \emph{Proceedings of the IEEE/CVF Conference on Computer Vision
  and Pattern Recognition}, pages 18134--18144, 2022.

\bibitem[Yang et~al.(2022)Yang, Liu, Zhu, Zhao, and Beroza]{yang2022toward}
Lei Yang, Xin Liu, Weiqiang Zhu, Liang Zhao, and Gregory~C Beroza.
\newblock Toward improved urban earthquake monitoring through
  deep-learning-based noise suppression.
\newblock \emph{Science advances}, 8\penalty0 (15):\penalty0 eabl3564, 2022.

\bibitem[Yang et~al.(2021)Yang, Hu, Zhang, and Liu]{yang2021simultaneous}
Shaobo Yang, Jing Hu, Haijiang Zhang, and Guiquan Liu.
\newblock Simultaneous earthquake detection on multiple stations via a
  convolutional neural network.
\newblock \emph{Seismological Research Letters}, 92\penalty0 (1):\penalty0
  246--260, 2021.

\bibitem[Yano et~al.(2021)Yano, Shiina, Kurata, Kato, Komaki, Sakai, and
  Hirata]{yano2021graph}
Keisuke Yano, Takahiro Shiina, Sumito Kurata, Aitaro Kato, Fumiyasu Komaki,
  Shin'ichi Sakai, and Naoshi Hirata.
\newblock Graph-partitioning based convolutional neural network for earthquake
  detection using a seismic array.
\newblock \emph{Journal of Geophysical Research: Solid Earth}, 126\penalty0
  (5):\penalty0 e2020JB020269, 2021.

\bibitem[Yu and Wang(2022)]{yu2022fastlink}
Ziye Yu and Weitao Wang.
\newblock Fastlink: a machine learning and gpu-based fast phase association
  method and its application to yangbi m s 6.4 aftershock sequences.
\newblock \emph{Geophysical Journal International}, 230\penalty0 (1):\penalty0
  673--683, 2022.

\bibitem[Zhang et~al.(2020)Zhang, Zhang, Yuan, Liu, Chen, and
  Li]{zhang2020locating}
Xiong Zhang, Jie Zhang, Congcong Yuan, Sen Liu, Zhibo Chen, and Weiping Li.
\newblock Locating induced earthquakes with a network of seismic stations in
  oklahoma via a deep learning method.
\newblock \emph{Scientific reports}, 10\penalty0 (1):\penalty0 1--12, 2020.

\bibitem[Zhang et~al.(2022)Zhang, Reichard-Flynn, Zhang, Hirn, and
  Lin]{zhang2022spatio}
Xitong Zhang, Will Reichard-Flynn, Miao Zhang, Matthew Hirn, and Youzuo Lin.
\newblock Spatio-temporal graph convolutional networks for earthquake source
  characterization.
\newblock \emph{Journal of Geophysical Research: Solid Earth}, page
  e2022JB024401, 2022.

\bibitem[Zhu et~al.(2022{\natexlab{a}})Zhu, Fang, Miao, Fan, Zhang, and
  Li]{zhu2022deep}
Zhu, Lihua Fang, Fajun Miao, Liping Fan, Ji~Zhang, and Zefeng Li.
\newblock Deep learning and transfer learning of earthquake and quarry-blast
  discrimination: Applications to southern california and eastern kentucky.
\newblock \emph{Authorea Preprints}, 2022{\natexlab{a}}.

\bibitem[Zhu et~al.(2022{\natexlab{b}})Zhu, Li, and Fang]{zhu2022ustc}
Jun Zhu, Zefeng Li, and Lihua Fang.
\newblock Ustc-pickers: a unified set of seismic phase pickers transfer learned
  for china.
\newblock \emph{Earthquake Science}, 36:\penalty0 1--11, 2022{\natexlab{b}}.

\bibitem[Zhu et~al.(2019{\natexlab{a}})Zhu, Peng, McClellan, Li, Yao, Li, and
  Fang]{zhu2019deep}
Lijun Zhu, Zhigang Peng, James McClellan, Chenyu Li, Dongdong Yao, Zefeng Li,
  and Lihua Fang.
\newblock Deep learning for seismic phase detection and picking in the
  aftershock zone of 2008 mw7. 9 wenchuan earthquake.
\newblock \emph{Physics of the Earth and Planetary Interiors}, 293:\penalty0
  106261, 2019{\natexlab{a}}.

\bibitem[Zhu and Beroza(2018)]{zhu19phasenet}
Weiqiang Zhu and Gregory~C Beroza.
\newblock Phasenet: a deep-neural-network-based seismic arrival-time picking
  method.
\newblock \emph{Geophysical Journal International}, 216\penalty0 (1):\penalty0
  261--273, 10 2018.

\bibitem[Zhu et~al.(2019{\natexlab{b}})Zhu, Mousavi, and
  Beroza]{zhu2019seismic}
Weiqiang Zhu, S~Mostafa Mousavi, and Gregory~C Beroza.
\newblock Seismic signal denoising and decomposition using deep neural
  networks.
\newblock \emph{IEEE Transactions on Geoscience and Remote Sensing},
  57\penalty0 (11):\penalty0 9476--9488, 2019{\natexlab{b}}.

\bibitem[Zhu et~al.(2022{\natexlab{c}})Zhu, Tai, Mousavi, Bailis, and
  Beroza]{zhu2022end}
Weiqiang Zhu, Kai~Sheng Tai, S~Mostafa Mousavi, Peter Bailis, and Gregory~C
  Beroza.
\newblock An end-to-end earthquake detection method for joint phase picking and
  association using deep learning.
\newblock \emph{Journal of Geophysical Research: Solid Earth}, 127\penalty0
  (3):\penalty0 e2021JB023283, 2022{\natexlab{c}}.

\end{thebibliography}

\bibliographystyle{plainnat}

\section*{Supplementary Information}\label{sec7}

\subsection*{Text S1. Evaluation matrix}

To evaluate the classification performance of our models, we employed Receiver Operating Characteristic curves (ROC) and Area Under the Curve (AUC). For the two-class classification problem (Figure S3), the model outputs probabilities for each class (earthquake or explosion). By applying a threshold, events in the test dataset are classified as earthquake or explosion. Therefore, we calculated True Positive (TP, correctly classified explosions), True Negative (TN, correctly classified earthquakes), False Positive (FP, earthquakes classified as explosions), and False Negative (FN, explosions classified as earthquakes) elements to derive metrics such as precision, recall, True Positive Rate (TPR), False Positive Rate (FPR), and f1-score:
\begin{equation}Precision=\frac{TP}{TP+FP}\end{equation}
\begin{equation}TPR = Recall=\frac{TP}{TP+FN}\end{equation}
\begin{equation}FPR=\frac{FP}{FP+TN}\end{equation}
\begin{equation}f1 =2 \times \frac{precision \times recall}{precision + recall}\end{equation}
By varying the threshold and calculating TPR and FPR at each point, we obtained the ROC curve for the binary classification problem (Figure S3). The AUC value represents the overall performance of the classifier across different thresholds and ranges between 0 and 1, where higher values indicate better performance.

For multi-class classification tasks, we converted them into multiple binary classification problems using the one-vs-rest approach. For instance, to generate Figure 2, we respectively treated earthquake, explosion and surface event as positive samples, and the other two categories as negative samples to calculate TP, TN, FP, and FN for each class. This allowed us to obtain ROC curves for each class in the multi-class classification problem. Additionally, we calculated the Macro-average, representing the arithmetic mean of the metrics across multiple classes.

\subsection*{Text S2. Training details}

During the training process, the model was optimized using the ADAM method. For different training procedures, we employed different loss function, learning rate, epochs and batch size (Table S3). All the model was selected based on the best performance on validation dataset.

\setcounter{figure}{0}
\renewcommand{\thefigure}{S\arabic{figure}}
\setcounter{table}{0}
\renewcommand{\thetable}{S\arabic{table}}

\begin{table*}[htb!]
\begin{center}
\begin{minipage}{\textwidth}
\caption{Data division of event classification and mechanism analyze tasks.}\label{tab1}
\begin{tabular*}{\textwidth}{@{\extracolsep{\fill}}ccccccc@{\extracolsep{\fill}}}
\toprule	

Task & Type & Train & Validation & Test & Total   \\
\midrule

&Earthquake  & 13170 & 1350 & 5480 & 20000 \\
&Explosion  & 10597 & 1030  & 4316 & 15943  \\
\multirow{-3}{*}{\makecell{Event \\ Classification}}
&Surface Event  & 6000 & 598  & 2314 & 8912  \\
%
\midrule
&Normal Fault & 654 & 54 & 97 & 805 \\
&Reverse Fault & 941 & 88  & 138 & 1167  \\
\multirow{-3}{*}{Mechanism}
&Strike Slip  & 705 & 58  & 100 & 863  \\

\bottomrule
\end{tabular*}
\footnotetext{}
\end{minipage}
\end{center}
\end{table*}

\begin{table}[htbp]
\begin{center}
\begin{minipage}{\textwidth}
\caption{Data division of event classification task for testing model's generalization ability.}\label{tab3}
\begin{tabular*}{\textwidth}{@{\extracolsep{\fill}}lcccccc@{\extracolsep{\fill}}}
\toprule%
Type & Earthquake & Blast \\
\midrule
Number of Events & 913 & 464  \\
Number of Waveforms & 18642 & 10435  \\
\bottomrule
\end{tabular*}
\footnotetext{}
\end{minipage}
\end{center}
\end{table}

\begin{table}[htbp]
\begin{center}
\begin{minipage}{\textwidth}
\caption{Training details in pre-training process and three downstream tasks.}\label{tab3}
\begin{tabular*}{\textwidth}{@{\extracolsep{\fill}}lcccccc@{\extracolsep{\fill}}}
\toprule%
Type & Pre-train & Classification & Location & Focal Mechanism \\
\midrule
Learning rate & 1e-4 & 1e-4 & 2e-4 & 1e-4 \\
Batch Size & 192 & 128 & 16 & 4 \\
Epochs & 100 & 100  & 200 & 200   \\
Loss function & 
Cross-entropy & 
Cross-entropy  & MSE & 
Cross-entropy   \\
\bottomrule
\end{tabular*}
\footnotetext{}
\end{minipage}
\end{center}
\end{table}


\begin{figure}[h!]%
\centering
\includegraphics[width=0.8\textwidth]{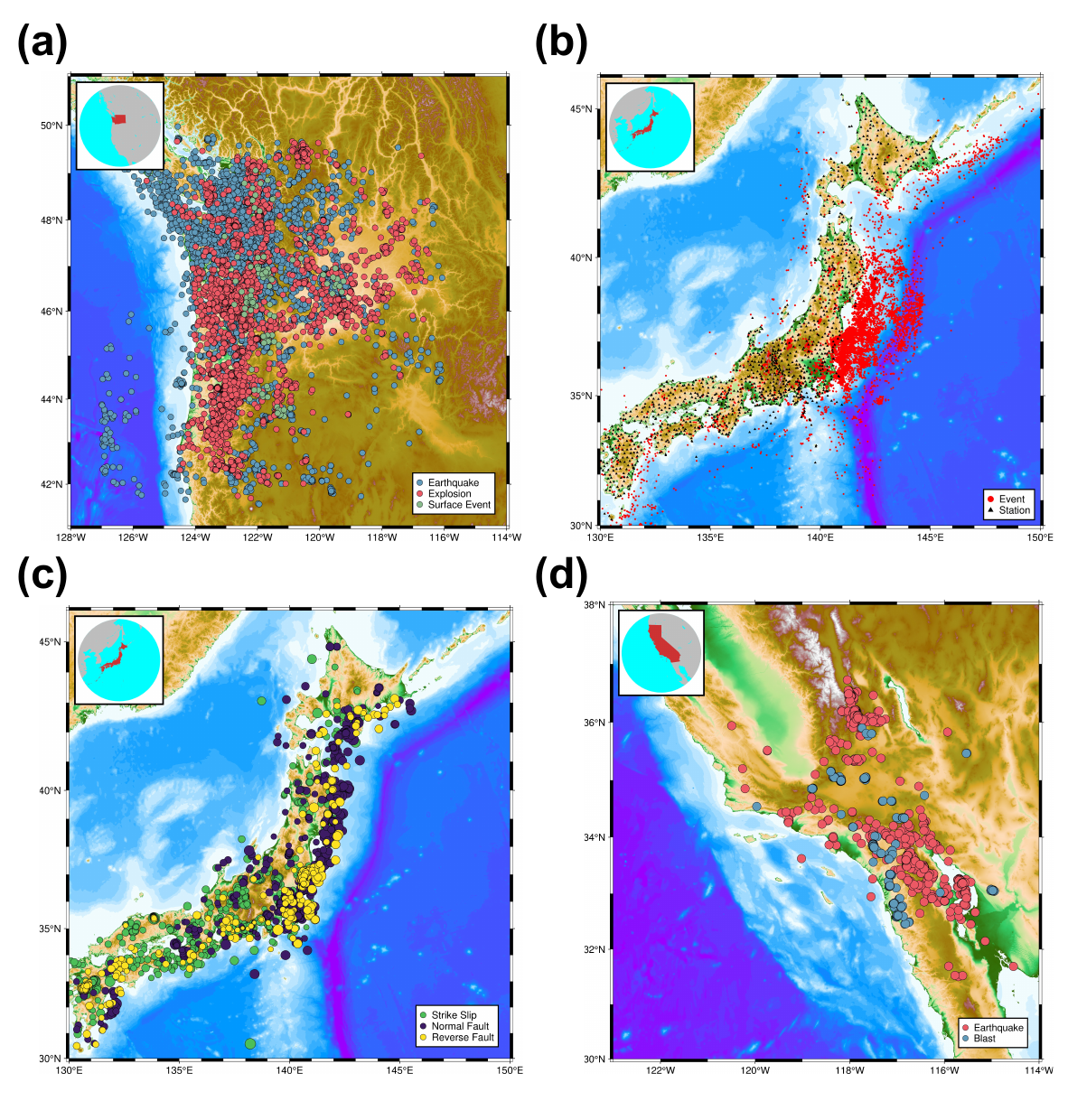}
\caption{Spatial distributions of events for different tasks. (a) Distribution of the events used for classification task from PNW datasets. The earthquake, explosion, and surface events are colored by blue, red, and green, respectively.  (b) Distribution of stations (black triangles) and event locations (red circles) used for the location task in our study, where the events occurred between January 1, 2011 and December 31, 2011. (c) Distribution of event locations used for the task of focal mechanism analysis in our study, where the green, dark blue and yellow circles respectively represent strike slip, normal fault and reverse fault. All the events occurred between 2011 and 2015. (b) Distribution of earthquakes (red circles) and explosions (blue circle) in southern California,  used to test the generalization ability of the models.}
\label{fig1}
\end{figure}

\begin{figure}[htbp]%
\centering
\includegraphics[width=\textwidth]{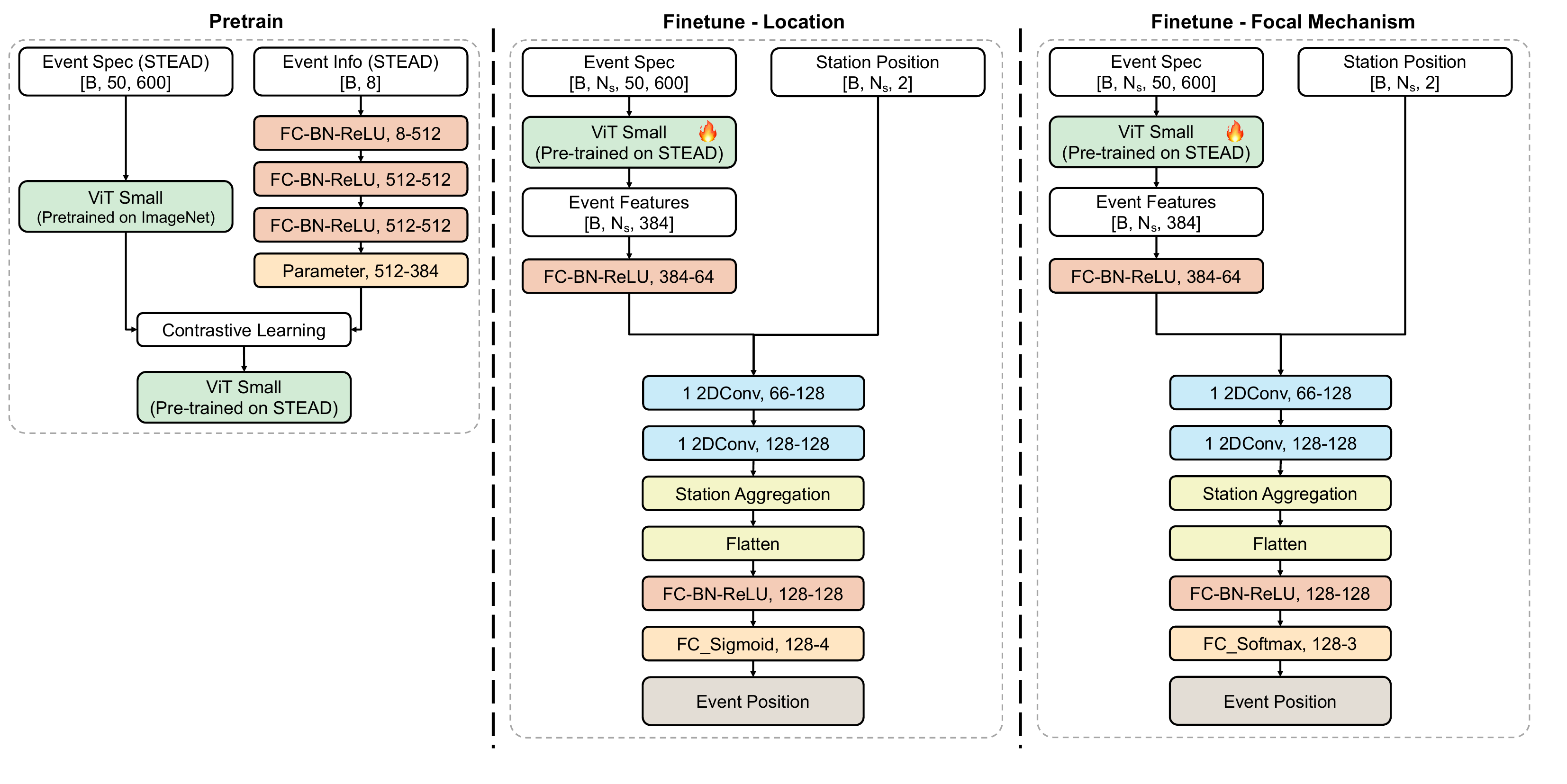}
\caption{While using the same pre-trained spectrum decoder (left panel) as in the event classification task (Figure 1), we designed different decoder networks for the location (middle panel) and focal mechanism (right panel) tasks.}
\label{fig3}
\end{figure}


\begin{figure}[htbp]%
\centering
\includegraphics[width=0.8\textwidth]{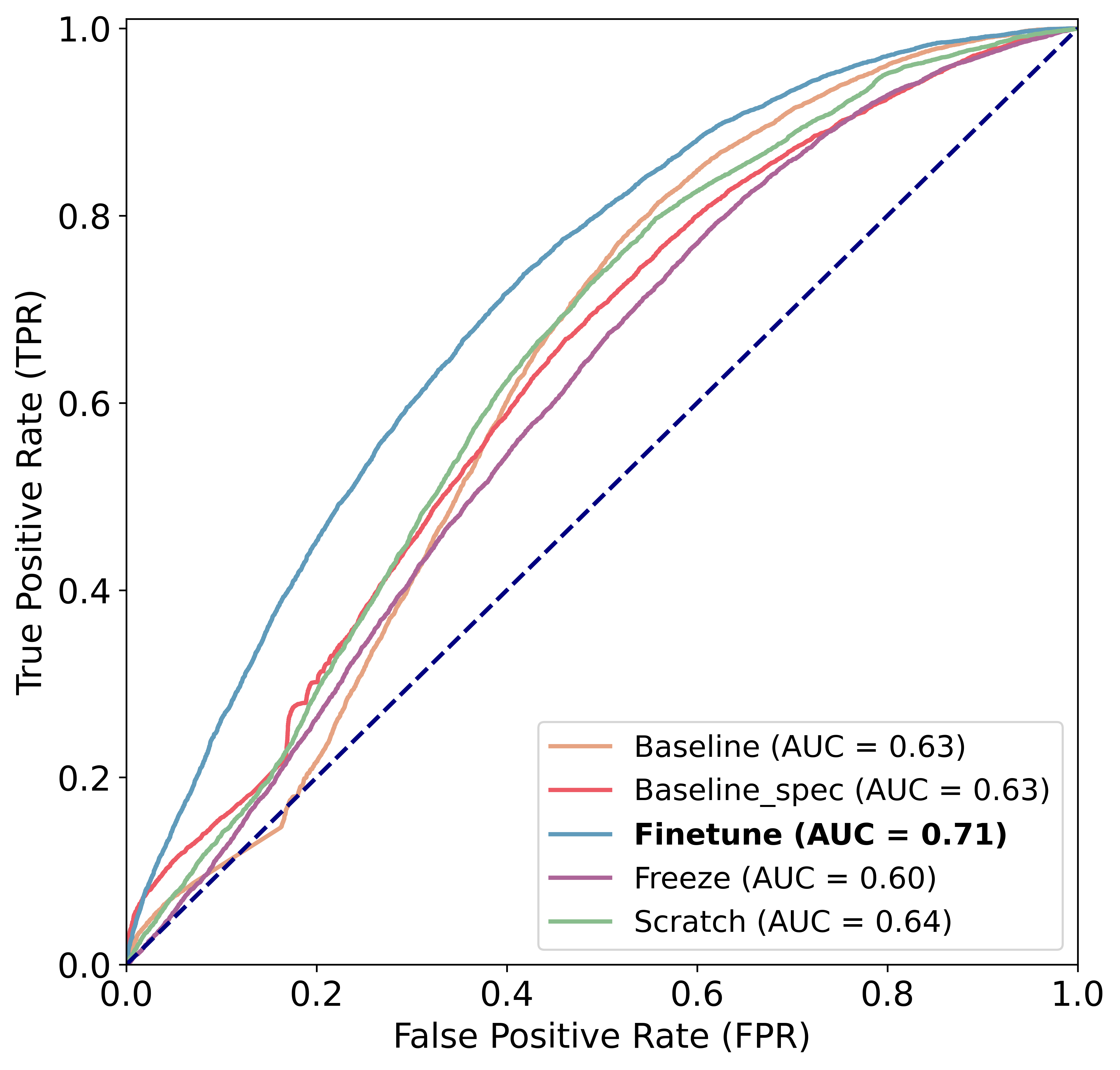}
\caption{The ROC curves of earthquake-explosion classification task for testing models' generalization ability. All of the models are initially trained on the PNW dataset and subsequently applied to the SCSN dataset. Given that it's a binary classification problem, we can generate the ROC curve for the earthquake-explosion classification task by adjusting the threshold and calculating the True Positive Rate (TPR) and False Positive Rate (FPR) at each threshold point.}
\label{fig8}
\end{figure} 

\begin{figure}[htbp]%
\centering
\includegraphics[width=0.8\textwidth]{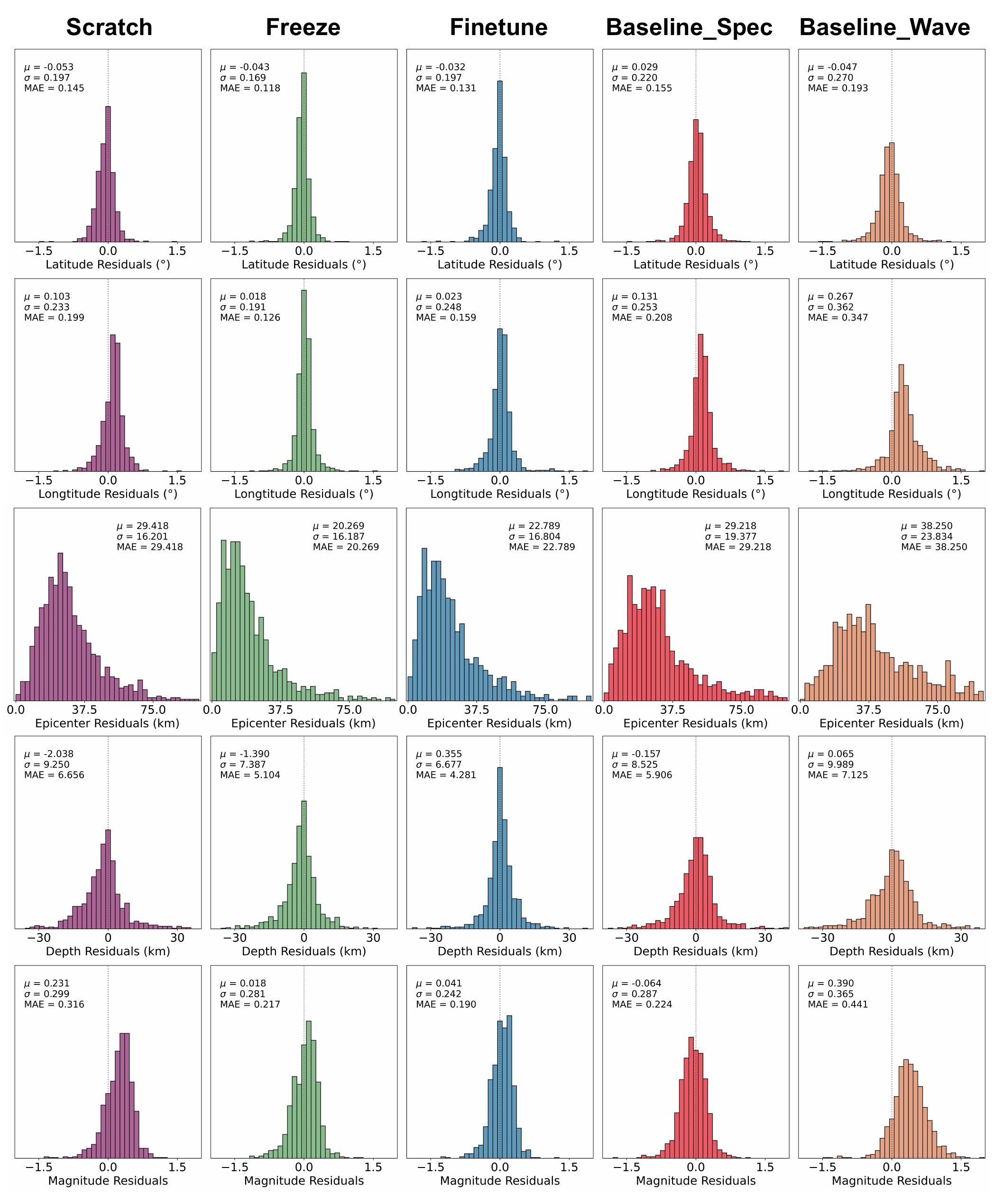}
\caption{Distributions of residuals in localization task. The first and second rows display latitude and longitude residuals, respectively. The third row shows epicenter residuals, calculated from latitude and longitude residuals. Notably, the frozen model and fine-tuned model exhibited the best performance. Additionally, the fourth and fifth rows illustrate depth and magnitude residuals, respectively. For depth and magnitude estimation, the fine-tune model demonstrated outstanding performance, as indicated by the narrowest distribution of residuals and the lowest mean absolute error (MAE).
}
\label{fig5}
\end{figure}


\begin{figure}[htbp]%
\centering
\includegraphics[width=\textwidth]{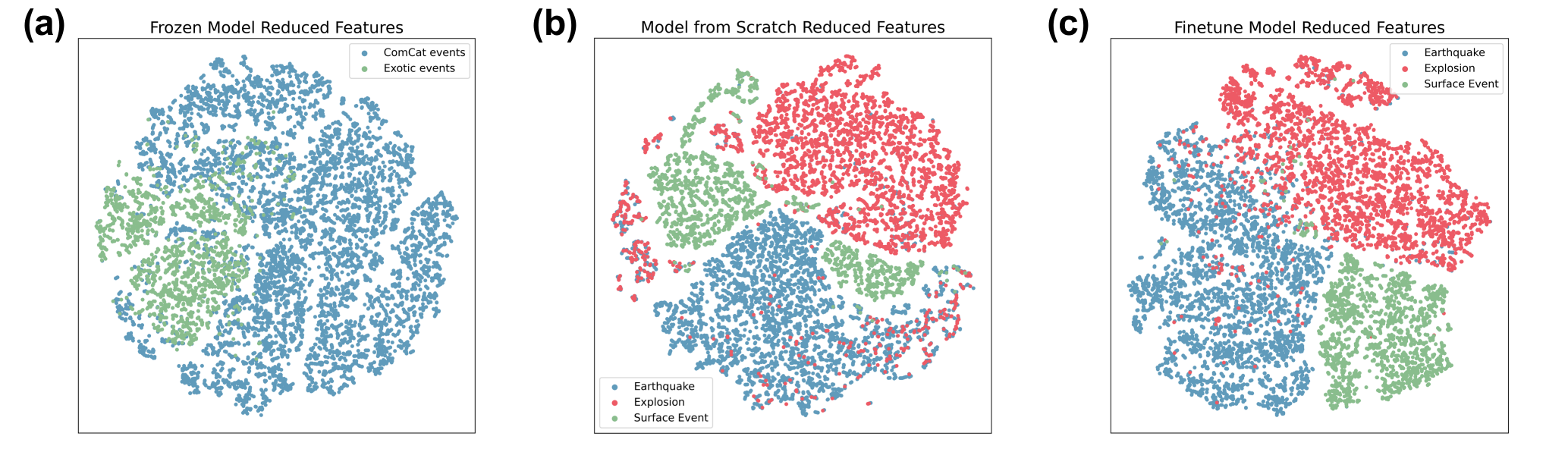}
\caption{The visualization of reduced features of three models. (a - c) The reduced features from the frozen, scratch, and fine-tuned models, which is calculated through t-SNE. 
Given our model's exclusive pre-training on earthquake events, the reduced features obtained from the frozen decoder are distinguishable primarily as `comcat events' (including earthquake and explosions) or `exotic events' (including surface events).
Moreover, in the fine-tuned model, the reduced features associated with earthquakes, explosions, and surface events are distinctly separated from each other, showing a much clearer distinction compared to those in the scratch model.
}\label{fig7}
\end{figure}


\end{document}